\documentclass[aps,pra,preprint,showpacs]{revtex4}
\usepackage{graphicx}
\usepackage{dcolumn}                 
\usepackage{times}
\usepackage{amsmath}

\newcommand*{\cent}[1]{\multicolumn{1}{c}{$#1$}}
\newcolumntype{w}[1]{D{.}{.}{#1}}
\newcolumntype{.}{D{x}{}{-1}}

\begin{document}
\preprint{Version 3.0}

\title{Born-Oppenheimer potential for H$_2$}

\author{Krzysztof Pachucki}
\email[]{krp@fuw.edu.pl}
\affiliation{Institute of Theoretical Physics, University of Warsaw,
             Ho\.{z}a 69, 00-681 Warsaw, Poland}

\begin{abstract}
The Born-Oppenheimer potential for the $^1\Sigma_g^+$ state of H$_2$ is obtained
in the range of 0.1 -- 20 au, using analytic formulas and recursion relations
for two-center two-electron integrals with exponential functions.
For small distances James-Coolidge basis is used, while for large
distances the Heitler-London functions with arbitrary polynomial in 
electron variables. In the whole range
of internuclear distance about $10^{-15}$ precision is achieved, as an example
at the equilibrium distance $r=1.4011$ au 
the Born-Oppenheimer potential amounts to $-1.174\,475\,931\,400\,216\,7(3)$.  
Results for the exchange energy verify 
the formula of Herring and Flicker [Phys. Rev. {\bf 134}, A362 (1964)]
for the large internuclear distance asymptotics.
The presented analytic approach to Slater integrals opens a window for the high precision 
calculations in an arbitrary diatomic molecule.
\end{abstract}

\pacs{31.15.ac, 31.50.Bc}
\maketitle

\section{Introduction}
In order to accurately obtain vibrational and rotational spectra of
molecules not only the nonrelativistic Born-Oppenheimer (BO) potential
has to be calculated, but also the adiabatic, nonadiabatic, relativistic and
quantum electrodynamics effects. An excellent agreement of theoretical results 
\cite{h2} with experimental values achieved for dissociation energies
of the hydrogen molecule H$_2$ \cite{h2_exp} and D$_2$ \cite{d2_exp}
indicates good understanding of all physical effects up to $10^{-8}$
precision level. Such a precision has not yet been achieved for 
molecular systems with more than two electrons. It is because
numerical calculations with explicitly correlated functions 
are very demanding. The commonly used explicitly correlated Gaussian functions
require global minimization of thousands of nonlinear parameters,
and the accuracy achieved for relativistic effects is difficult to estimate,
due to the improper analytic properties of these functions.
The largest system considered so far is the helium dimer, where about
$10^{-5}$ precision is achieved for the interaction energy \cite{he2}. 

Herewith, we apply explicitly correlated exponential functions
(for a recent review see Ref. \cite{correlated}),
in order to increase numerical accuracy obtained so far for two electrons
diatomic systems. The use of exponential functions 
may allow for a significant improvement not only in
rotational and vibrational energies, but also in other important properties
like magnetic shielding, spin-rotational or spin-spin coupling constants.
It is because evaluation of relativistic effects is very sensitive to 
correct analytic properties of the basis functions, a good example being
the relativistic correction to the magnetic shielding \cite{rudzins}. 

When $10^{-10}$ precision is achieved for vibrational transitions,
the electron-proton mass ratio can be determined with better accuracy,
than it is known presently \cite{nist}. Moreover, 
any nonelectromagnetic long-range interactions between nuclei, 
which cannot in principle be
excluded, may be visible not only in vibrational spectra, but also 
in the spin interaction between nuclei \cite{spindep}, for example in HD molecule.
The electromagnetic spin interaction is extremely small, of order $m^3/(m_p\,m_d)\,\alpha^6$,
about 43 Hz for the ground molecular state \cite{chemrev}, and any deviation
between experimental values and theoretical predictions will signal
existence of a nonelectromagnetic long-range interactions between nuclei.
 
In this work we present an effective computational method with the use of
exponential functions and numerical results for 
the Born-Oppenheimer energy of the H$_2$ molecule. 
We demonstrate that  $10^{-15}$ numerical precision can be achieved for 
BO energy with about 22 000 basis functions
using the analytic formulae and recursion relations for Slater integrals, 
which have been obtained in our previous work in \cite{rec_h2}.
The obtained BO potential is at least two orders of magnitude more accurate,
than anything published so far, and at the equilibrium distance
our calculations confirm the most accurate so far result of Cencek and
Szalewicz in Ref. \cite{cencek} well within their uncertainties. 
Our results for large internuclear distances verify 
the asymptotic formula of Herring and Flicker \cite{herring}, 
although we observe significant contributions from the off leading terms.
Finally, we analyze  possibility of 
the extension of this method to more than two-electron systems.

\section{ Short range of internuclear distances}
The Schr\"odinger equation for H$_2$ molecule
in the Born-Oppenheimer approximation is
\begin{equation}
H\phi(\vec r_1\,,\vec r_2) = E(r)\phi(\vec r_1\,,\vec r_2),
\end{equation}
where
\begin{equation}
H =
-\frac{\nabla_1^2}{2}-\frac{\nabla_2^2}{2}-\frac{1}{r_{1A}}-\frac{1}{r_{1B}}
-\frac{1}{r_{2A}}-\frac{1}{r_{2B}} +\frac{1}{r_{12}}+\frac{1}{r}.
\end{equation}
The most efficient basis set to solve this equation is the one
introduced by Ko\l os and Wolniewicz in \cite{kw_basis}.
However at small nuclear distances $r\equiv r_{AB}$, 
we can use much simpler James-Coolidge basis \cite{james_cool}, 
namely the functions of the form
\begin{eqnarray}
\phi &=& \sum_{\{n\}}\,c_{\{n\}}
(1+P_{AB})\,(1+P_{12})\,e^{-\alpha\,(r_{1A}+r_{1B})}\,e^{-\alpha\,(r_{2A}+r_{2B})}\,r_{12}^{n_1}\nonumber \\ &&
(r_{1A}-r_{1B})^{n_2}\,(r_{2A}-r_{2B})^{n_3}\,(r_{1A}+r_{1B})^{n_4}\,(r_{2A}+r_{2B})^{n_5} 
\end{eqnarray}
such that
\begin{equation}
\sum_{i=1}^5 n_i \leq \Omega
\end{equation}
with $\Omega=3, \ldots 20$. This basis have recently been used
by Sims and Hagstrom \cite{SH06} for $10^{-12}$ precision
calculation of BO potential in the range $r=0.4 - 6$ au, by a traditional
way of numerical evaluation of the corresponding two-center integrals.
Here, using Ref. \cite{rec_h2} we derive analytic expression for all these integrals
up to $\Omega=20$. They have very simple form and involve only exponential integral (Ei), 
exponential and logarithmic functions.  
In this derivation we make use of a differential equation, which 
is satisfied by the most general two-center two-electron Slater integral, 
see Ref. \cite{rec_h2}.

For the presentation of analytic formulas we consider an integral
\begin{eqnarray}
f(n_1,n_2,n_3,n_4,n_5;r,u,w) &=& \int \frac{d^3 r_1}{4\,\pi}\,\int \frac{d^3 r_2}{4\,\pi}\,
\frac{e^{-u\,r_{1A}}}{r_{1A}}\,
\frac{e^{-u\,r_{1B}}}{r_{1B}}\,
\frac{e^{-w\,r_{2A}}}{r_{2A}}\,
\frac{e^{-w\,r_{2B}}}{r_{2B}}\,\frac{r}{r_{12}^{1-n_1}}\nonumber \\ &&
(r_{1A}-r_{1B})^{n_2}\,(r_{2A}-r_{2B})^{n_3}\,(r_{1A}+r_{1B})^{n_4}\,(r_{2A}+r_{2B})^{n_5}
\end{eqnarray}
at the internuclear distance $r=1$, since arbitrary distances can be obtained
from
the relation 
\begin{equation}
f(n_1,n_2,n_3,n_4,n_5;r,u,w) = f(n_1,n_2,n_3,n_4,n_5,1,r\,u,r\,w)\,r^{2+n_1+n_2+n_3+n_4+n_5}
\end{equation}
Positive powers of $n_4$ and $n_5$ can be obtained by differentiation with respect to
$u$ and $w$ correspondingly, so without loss of generality one can assume $n_4=n_5=0$.
Let us introduce auxiliary functions $y_i$ defined by
\begin{eqnarray}
y_0 &=& e^{-(u+w)}\\
y_1 &=& {\rm Ei}(-2\,w)\,e^{w-u} + {\rm Ei}(-2\,u)\,e^{u-w}\\
y_2 &=& {\rm Ei}(-2\,w)\,e^{w-u} - {\rm Ei}(-2\,u)\,e^{u-w}\\
y_3 &=& {\rm Ei}\bigl(-2\,(u+w)\bigr)\,e^{u+w} +\biggl[\ln\Bigl(\frac{2\,u\,w}{u+w}\Bigr)+\gamma\biggr]\,e^{-(u+w)}\\
y_4 &=& {\rm Ei}\bigl(-2\,(u+w)\bigr)\,e^{u+w} -\biggl[\ln\Bigl(\frac{2\,u\,w}{u+w}\Bigr)+\gamma\biggr]\,e^{-(u+w)}
\end{eqnarray}
Then, using the equation (69) from \cite{rec_h2}, 
all nonvanishing integrals $f(n_1,n_2,n_3) \equiv f(n_1,n_2,n_3,0,0;1,u,w)$ 
with $\sum_i n_i\leq 4$ are the following:
\begin{eqnarray}
f(0,0,0) &=&\frac{y_3-y_1}{4\,u\,w}\\[1ex]
f(1,0,0) &=&\frac{y_0}{4\,u\,w} \\[1ex]
f(2,0,0) &=&\frac{y_0\,(2\,u\,w+u+w)+y_2\,(u-w)-y_4\,(u+w)}{24\,u^2\,w^2}
                +\frac{(y_3-y_1) \left(u^2+w^2\right)}{24\,u^3\,w^3}\\[1ex]
f(0,2,0) &=& \frac{y_3-y_1}{12\,u\,w}\\[1ex]
f(0,1,1) &=&-\frac{y_0}{6\,u\,w} + \frac{y_1\,(u\,w -1) - y_2\,(u-w) + y_3\,(u\,w+1) - y_4\,(u+w)}{12\,u^2\,w^2}\\[1ex]
f(3,0,0) &=&\frac{y_0 \left(u^2\,w^2+3\,u\,w\,(u+w)+3\,(u^2+w^2)\right)}{24\,u^3\,w^3}\\[1ex]
f(1,2,0) &=& \frac{y_0}{12\,u\,w}
\end{eqnarray}
\begin{eqnarray}
f(4,0,0) &=&\frac{y_0 \left(2 u^3 w^3+6 u^3
    w^2+9 u^3 w+3 u^3+6 u^2 w^3+4 u^2 w^2+6 u^2 w+9 u w^3+6 u w^2+3
    w^3\right)}{80 u^4 w^4}
    \nonumber \\&&+\frac{y_2 (u-w) \left(3 u^2+u w+3 w^2\right)}{40 u^4 w^4}-\frac{y_4 (u+w)
    \left(3 u^2-u w+3 w^2\right)}{40 u^4 w^4}
    \nonumber\\&&
+\frac{(y_3-y_1)\left(u^4 w^2+3 u^4+u^2 w^4+2 u^2 w^2+3 w^4\right)}{40 u^5 w^5}
+\frac{y_3+y_1}{20 u^2 w^2}\\ [1ex]
f(2,2,0) &=&  \frac{y_2 (u-w) (u w+2)}{120 u^3 w^3}-\frac{y_4 (u+w) (u w-2)}{120 u^3
    w^3}+\frac{y_0 (4 u w+3 u+3 w+4)}{120 u^2 w^2}
    \nonumber \\&&-\frac{y_1 \left(u^2+2 u
    w+w^2-2\right)}{120 u^3 w^3}+\frac{y_3 \left(u^2-2 u w+w^2-2\right)}{120
    u^3 w^3} \\[1ex]
f(2,1,1) &=& -\frac{y_0 \left(2
    u^2 w^2+3 u^2 w+6 u^2+3 u w^2+6 w^2\right)}{120 u^3 w^3}
    \nonumber\\&&-\frac{y_2 (u-w) \left(u^2 w^2+3 u^2+3 w^2\right)}{120 u^4 w^4}-\frac{y_4
    (u+w) \left(u^2 w^2+3 u^2+3 w^2\right)}{120 u^4 w^4}
    \\&&+\frac{(y_3+y_1)
    \left(u^2+w^2\right)}{40 u^3
    w^3}+\frac{(y_3-y_1) \left(2 u^2 w^2+3 u^2+3 w^2\right)}{120
    u^4 w^4} \nonumber\\[1ex]
f(0,4,0) &=&\frac{y_3-y_1}{20\,u\,w} \\[1ex]
f(0,3,1) &=&  \frac{y_1 (u w-1)}{20 u^2 w^2}-\frac{y_2 (u-w)}{20 u^2 w^2}+\frac{y_3 (u
    w+1)}{20 u^2 w^2}-\frac{y_4 (u+w)}{20 u^2 w^2}-\frac{y_0}{10 u w}\\[1ex]
f(0,2,2) &=& \frac{y_2 (u-w) (u w-3)}{15 u^3 w^3}-\frac{y_4 (u+w) (u w+3)}{15 u^3
    w^3}-\frac{y_0 (u w+2 u+2 w+6)}{15 u^2 w^2}
    \nonumber\\&&+\frac{y_3+y_1}{2 u^2 w^2}+\frac{(y_3-y_1) \left(3 u^2 w^2+4
    u^2+4 w^2+12\right)}{60 u^3 w^3}
\end{eqnarray}
Integrals with higher powers of $n_i$ are similar, with simple polynomial
form in $1/u$ and $1/w$. We have tabulated all integrals, such that $\sum_i
n_i \leq 45$ what is sufficient for matrix elements involving exponential
functions with maximum value of $\Omega=20$.
 
\section{ Long range of internuclear distances}
For the large internuclear distance the James-Coolidge basis is 
not appropriate as it does not include the Heitler-London function.
Instead, we employ generalized Heitler-London functions, which are 
the product of a Heitler London function with
an arbitrary polynomial in all electron distances, 
\begin{eqnarray}
\phi &=& \sum_{\{n\}}\,c_{\{n\}}
(1+P_{AB})\,(1+P_{12})\,e^{-(r_{1A}+r_{2B})}
\,r_{12}^{n_1}\,r_{1A}^{n_2}\,r_{1B}^{n_3}\,r_{2A}^{n_4}\,r_{2B}^{n_5}
\end{eqnarray}
such that
\begin{equation}
\sum_{i=1}^5 n_i \leq \Omega,
\end{equation}
and call them the explicitly correlated asymptotic (ECA) basis.
Matrix elements of the nonrelativistic Hamiltonian
can be expressed in terms of direct integrals of the form 
\begin{equation}
f(n_1,n_2,n_3,n_4,n_5;r,x) = \int \frac{d^3 r_1}{4\,\pi}\,\int \frac{d^3 r_2}{4\,\pi}\,
\frac{r}{r_{12}^{1-n_1}}\,
\frac{e^{-x\,r_{1A}}}{r_{1A}^{1-n_2}}\,
\frac{1}{r_{1B}^{1-n_3}}\,
\frac{1}{r_{2A}^{1-n_4}}\,
\frac{e^{-x\,r_{2B}}}{r_{2B}^{1-n_5}}.
\end{equation}
with nonnegative integers $n_i$, and exchange integrals
which coincide with that from the previous section.
The direct integrals are calculated as follows.
When all $n_i=0$ the so called master integral is given by \cite{rec_h2}
\begin{eqnarray}
f(r,x) &=& \int \frac{d^3 r_1}{4\,\pi}\,\int \frac{d^3 r_2}{4\,\pi}\,
\frac{e^{-x\,r_{1A}}}{r_{1A}}\,\frac{e^{-x\,r_{2B}}}{r_{2B}}\,
\frac{1}{r_{1B}}\,\frac{1}{r_{2A}}\,\frac{r}{r_{12}}\\
&=& \frac{1}{x}\,\biggl[I_0(\sqrt{2}\,x\,r)\int_r^\infty dr'\,F(x\,r')\,K_0(\sqrt{2}\,x\,r')
    +K_0(\sqrt{2}\,x\,r)\int_0^rdr'\,F(x\,r')\,I_0(\sqrt{2}\,x\,r')\biggr]\nonumber
\end{eqnarray}
and its first derivative with respect to $r$
\begin{equation}
f'(r,x) = \sqrt{2}\,\biggl[I_1(\sqrt{2}\,x\,r)\int_r^\infty dr'\,F(x\,r')\,K_0(\sqrt{2}\,x\,r')
    -K_1(\sqrt{2}\,x\,r)\int_0^rdr'\,F(x\,r')\,I_0(\sqrt{2}\,x\,r')\biggr]
\end{equation}
where
\begin{eqnarray}
F(x) &=& -e^{x}\,\,[{\rm Ei}(-2\,x)+{\rm Ei}(-x)]+
          e^{-x}\,[{\rm Ei}(x)+2\,{\rm Ei}(-x)-\gamma-\ln(2\,x)].
\end{eqnarray}
This one dimensional integration can easily be obtained with the sufficient
accuracy and for example at $r=6,\; x=2$ the value of $f$ is
\begin{equation}
       f(6,2)  =  0.001\,790\,194\,708\,681\,916\,168\,871\,495\,878\,339\,876
\end{equation}
All integrals with  higher powers of electron distances can be obtained from recursion
relations, which were derived in Ref. \cite{rec_h2}. Since  
\begin{equation}
f(n_1,n_2,n_3,n_4,n_5;r,x) = 
f(n_1,n_2,n_3,n_4,n_5,1,r\,x)\,r^{2+n_1+n_2+n_3+n_4+n_5}
\end{equation}
we can present formulae at $r=1$. Let us introduce auxiliary functions 
\begin{eqnarray}
y_0 &=& e^{-x}\\
y_1 &=& [{\rm Ei}(x) - \ln(2\,x)-\gamma]\, e^{-x} - [{\rm Ei}(-2\,x) - {\rm Ei}(-x)]\, e^x\\
y_2 &=& [{\rm Ei}(x) - \ln(2\,x)-\gamma\,] e^{-x} + [{\rm Ei}(-2\,x) - {\rm Ei}(-x)]\, e^x\\
y_3 &=& [2\,{\rm Ei}(-x) + {\rm Ei}(x) - \ln(2\,x)-\gamma]\, e^{-x} + 
        [{\rm Ei}(-2\,x) + {\rm Ei}(-x)]\, e^x\\
y_4 &=& [2\,{\rm Ei}(-x) + {\rm Ei}(x) - \ln(2\,x)-\gamma]\, e^{-x} - 
        [{\rm Ei}(-2\,x) + {\rm Ei}(-x)]\, e^x
\end{eqnarray}
then all integrals $f(n_1,n_2,n_3,n_4,n_5; x)$ with $\sum_i n_i\leq 2$ are the following
\begin{eqnarray}
f(0, 0, 0, 0, 0) &=& f \\
f(0, 0, 1, 0, 0) &=& \frac{f}{x} - \frac{f'}{2\,x} \\
f(0, 1, 0, 0, 0) &=& \frac{y_2}{2 x^3} \\
f(1, 0, 0, 0, 0) &=& \frac{(1-y_0)^2}{x^4} \\
f(0, 0, 0, 2, 0) &=& \frac{y_0^2-1}{2 x^4}+\frac{y_0}{2
    x^3}+\left(\frac{2}{x^2}+\frac{1}{2}\right) f-\frac{f'}{x^2} \\
f(0, 0, 1, 1, 0) &=& -\frac{y_0^2-1}{2 x^4}-\frac{y_4}{2 x^3}-\frac{y_0}{2
    x^3}+\left(\frac{1}{x^2}+\frac{1}{2}\right) f-\frac{3 f'}{2 x^2} \\
f(0, 0, 2, 0, 0) &=& \frac{y_0^2-1}{2 x^4}+\frac{y_0}{2
    x^3}+\left(\frac{2}{x^2}+\frac{1}{2}\right) f-\frac{f'}{x^2} \\
f(0, 2, 0, 0, 0) &=& \frac{y_0^2}{x^4}+\frac{y_4}{2 x^3}+\frac{f}{x^2}+\frac{f'}{2
    x^2}+\left(\frac{1}{2 x^3}-\frac{1}{x^4}\right) y_0 \\
f(2, 0, 0, 0, 0) &=& -\frac{y_4}{x^5}-\frac{2 y_0}{x^5}-\frac{y_3}{2
    x^4}+\frac{1}{x^4}+\frac{2
    f}{x^2}+\left(\frac{2}{x^5}+\frac{1}{x^4}\right)
    y_0^2-\left(\frac{1}{x^4}+\frac{1}{2 x^2}\right) f' \\ 
f(0, 1, 0, 0, 1) &=& \frac{1}{x^4}-\left(\frac{1}{x^4}+\frac{1}{2 x^3}\right) y_0 \\
f(0, 1, 0, 1, 0) &=& \frac{y_2}{2 x^4}+\frac{y_1}{2 x^3}+\frac{y_0}{2 x^3} \\
f(0, 1, 1, 0, 0) &=& \frac{y_2}{x^4}-\frac{y_0}{2 x^3} \\
f(1, 0, 1, 0, 0) &=&  \frac{2}{x^5}-\left(\frac{4}{x^5}+\frac{1}{x^4}\right)
    y_0+\left(\frac{2}{x^5}+\frac{1}{x^4}\right) y_0^2 \\
f(1, 1, 0, 0, 0) &=& \frac{1-y_0}{x^4}
\end{eqnarray}
Integrals with higher powers $n_i$ are of analogous form,
they are  all  linear combinations of $f$, $f'$, $y_0^2$ and $y_i$
with coefficients being polynomials of $1/x$.
We have generated a table of integrals with $\sum_i n_i \leq 37$,
which corresponds to a maximum value of $\Omega=16$. It is less than $\Omega=20$ in the
case of James-Coolidge functions, but anyway, requires the use of octuple
precision arithmetics due to near linear dependence of these basis functions.

\section{Numerical results}
The matrix elements of the nonrelativistic Hamiltonian  between exponential
functions are obtained as described by Ko\l os and Roothan in \cite{kolroth}. 
The resulting expression is a linear combination of various Slater integrals 
which are calculated using analytic formulas as presented in the previous
sections. This evaluation is fast and accurate, thus allowing
the use of a large number of basis functions.

Eigenvalues of the Hamiltonian matrix are obtained 
by inverse iteration method for various length of the basis set.
Following Sims and Hagstrom \cite{SH06} we use double basis set $(\Omega, \Omega-2)$
with two different nonlinear parameters, which were obtained by minimization
of energy at $\Omega=12$. For the calculations with basis functions
up to $\Omega=20$, the second nonlinear parameter is additionally multiplied
by 1.5 in order to improve the numerical stability for the large values of $\Omega$. 
Numerical calculations with James-Coolidge functions are performed in general
using quadruple precision arithmetics, and for checking the numerical accuracy,
the one point at $r=1.4011$ au is calculated using the octuple precision.
We observe a significant loss of digits for large values of $\Omega$. The quadruple
precision arithmetics was not always sufficient for the whole range of internuclear
distances. In many cases values with $\Omega=20$, and sometimes even $\Omega=19$
had to be disregarded due to numerical instabilities in the inverse
iteration procedure. Numerical results at $r=1.4011$ and $r=6$ au for various sizes 
of basis length are presented in Table \ref{table1}.
\begin{table}[htb]\renewcommand{\arraystretch}{0.85}
\caption{\label{table1} Numerical values for BO energy 
obtained with James-Coolidge functions,
double basis set $(\Omega, \Omega-2)$ with nonlinear parameters:
$\alpha_1 = 0.9650,\; \alpha_2 = 4.6716$ for $r=1.4011$ and
$\alpha_1 = 0.57050,\; \alpha_2 = 2.36925$ for $r=6.0$}
\begin{ruledtabular}
  \begin{tabular}{rr..}
 $\Omega$ & N & \cent{r=1.4011} & \cent{r=6.0} \\ \hline
  3 &23&  -1.173\,189\,x743\,241\,832\,33         &  -0.969\,640\,x616\,224\,699\,7   \\
  4 &51&  -1.174\,345\,x859\,790\,847\,31         &  -0.995\,618\,x908\,838\,543\,1   \\
  5 &98&  -1.174\,463\,x692\,797\,208\,81         &  -1.000\,157\,x746\,708\,230\,0   \\
  6 &180&  -1.174\,474\,x907\,247\,221\,29        &  -1.000\,781\,x343\,551\,690\,1   \\
  7 &306&  -1.174\,475\,x841\,117\,269\,47        &  -1.000\,831\,x730\,125\,255\,5   \\
  8 &501&  -1.174\,475\,x923\,062\,032\,59        &  -1.000\,835\,x509\,296\,673\,5   \\
  9 &781&  -1.174\,475\,x930\,460\,077\,70        &  -1.000\,835\,x696\,691\,923\,9   \\
  10 &1182& -1.174\,475\,x931\,266\,449\,18       &  -1.000\,835\,x707\,165\,465\,4   \\
  11 &1729& -1.174\,475\,x931\,378\,701\,09       &  -1.000\,835\,x707\,596\,974\,8   \\
  12 &2471& -1.174\,475\,x931\,396\,395\,07       &  -1.000\,835\,x707\,643\,241\,4   \\
  13 &3444& -1.174\,475\,x931\,399\,453\,43       &  -1.000\,835\,x707\,651\,449\,9   \\
  14 &4712& -1.174\,475\,x931\,400\,035\,47       &  -1.000\,835\,x707\,653\,956\,0   \\
  15 &6324& -1.174\,475\,x931\,400\,162\,72       &  -1.000\,835\,x707\,654\,760\,4   \\
  16 &8361& -1.174\,475\,x931\,400\,197\,06       &  -1.000\,835\,x707\,655\,025\,5   \\
  17 &10887& -1.174\,475\,x931\,400\,208\,48      &  -1.000\,835\,x707\,655\,120\,8   \\
  18 &14002& -1.174\,475\,x931\,400\,213\,00      &  -1.000\,835\,x707\,655\,156\,0   \\
  19 &17787& -1.174\,475\,x931\,400\,215\,01      &  -1.000\,835\,x707\,655\,168\,9   \\
  20 &22363& -1.174\,475\,x931\,400\,215\,99      &  -1.000\,835\,x707\,655\,175\,9   \\[1ex]
$\infty$&&   -1.174\,475\,x931\,400\,216\,7(3)    &  -1.000\,835\,x707\,655\,180\,4(22) \\
$\infty$&Cencek \cite{cencek}&  -1.174\,475\,x931\,400\,21(6)   
  \end{tabular}
\end{ruledtabular}
\end{table}
The most accurate  result obtained at the equilibrium distance $r=1.4011$ au 
is compared in Table \ref{table2} to all the previous results obtained so far in the literature. 
\begin{table}[htb]\renewcommand{\arraystretch}{0.85}
\caption{\label{table2} Born-Oppenheimer potential for the H$_2$ molecule
  with different types of wave functions at $r=1.4011$ au}
\begin{ruledtabular}
  \begin{tabular}{lcr.}
 Authors                                     & type of w.f. &  N  & {\rm energy}                \\ \hline\\[-5pt]
 1933\; James and Coolidge \cite{james_cool} & JC           &  5  & -1.173\,x5                  \\
 1960\; Ko\l os and Roothan \cite{kolroth}   & JC           &  50 & -1.174\,x448                \\
 1968\; Ko\l os and Wolniewicz \cite{kolwol} & KW           & 100 & -1.174\,x474\,983      \\
 1994\; Rychlewski, Cencek, Komasa \cite{rych}& ECG         & 700 & -1.174\,x475\,931           \\
 1995\; Wolniewicz \cite{wol}                & KW           & 883 & -1.174\,x475\,930\,742      \\
 2006\; Sims and Hagstrom  \cite{SH06}       & JC           &7034 & -1.174\,x475\,931\,399\,84  \\
 2007\; Nakatsuji {\em et al.}\cite{nakatsuji}& ICI         &6776 & -1.174\,x475\,931\,400\,027 \\
 2008\; Cencek, Szalewicz \cite{cencek}      & ECG          &4800 & -1.174\,x475\,931\,400\,135 \\
 2010\; this work                            & JC           &22363& -1.174\,x475\,931\,400\,215\,99  
  \end{tabular}
\end{ruledtabular}
\end{table}

In performing extrapolation,  we observe the exponential $e^{-\beta\,\Omega}$
convergence. In other words the $\log$ of differences
in energies for subsequent values of $\Omega$ fits well to the linear
function. This allows for  a simple and reliable extrapolation to infinity.
For example at the equilibrium distance the parameter $\beta$ is about $0.9$,
and convergence in the James-Coolidge
basis preserve its exponential behavior for all the distances.

\begin{figure}[!htb]
\centering
\includegraphics[width=15cm]{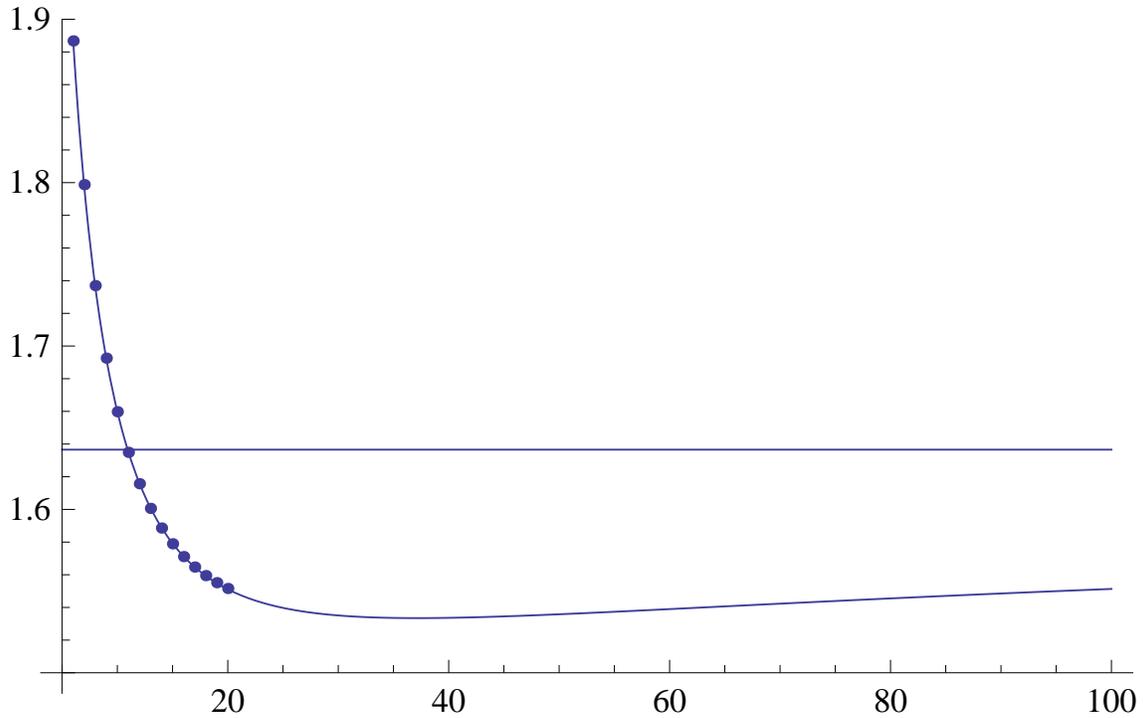}
\caption{\label{figure1} Rescaled exchange energy $e^{2\,r}\,r^{-5/2}\,\Delta E$ 
fitted to numerical points in Table \ref{table4} in the range $r=6 - 20$ au, 
assuming Herring and Flicker \cite{herring} asymptotics, straight line.}
\end{figure}

The calculations for the internuclear distances $r=6 - 20$ au.
are performed in ECA basis set using octuple precision arithmetics up to
$\Omega=16$. This basis is probably the most effective one at large distances.
We also observe exponential convergence for BO and even exchange energies, 
what makes quite simple the extrapolation to infinity,
see Table \ref{eca} with detailed results for various length of the basis set.
\begin{table}[htb]\renewcommand{\arraystretch}{0.85}
\caption{\label{eca} Numerical values for BO and exchange energies 
obtained with ECA functions at $r=12$ au}
\begin{ruledtabular}
  \begin{tabular}{rr..}
 $\Omega$ & N & E(^1\Sigma_g^+)            & \cent{e^{2\,r}\,\Delta E} \\ \hline
  3  &32  &-1.000\,002\,x515\,756\,9814    & 720.758\,x075 \\
  4  &70  &-1.000\,002\,x541\,561\,9799    & 767.006\,x890 \\
  5  &136 &-1.000\,002\,x545\,005\,4934    & 791.784\,x126 \\
  6  &246 &-1.000\,002\,x545\,635\,6130    & 800.649\,x925 \\
  7  &416 &-1.000\,002\,x545\,838\,2275    & 803.989\,x841 \\
  8  &671 &-1.000\,002\,x545\,922\,5168    & 805.314\,x087 \\
  9  &1036&-1.000\,002\,x545\,955\,3661    & 805.779\,x212 \\
  10 &1547&-1.000\,002\,x545\,965\,8295    & 805.916\,x877 \\
  11 &2240&-1.000\,002\,x545\,968\,6694    & 805.953\,x991 \\
  12 &3164&-1.000\,002\,x545\,969\,3389    & 805.963\,x087 \\
  13 &4368&-1.000\,002\,x545\,969\,4859    & 805.965\,x246 \\
  14 &5916&-1.000\,002\,x545\,969\,5175    & 805.965\,x766 \\
  15 &7872&-1.000\,002\,x545\,969\,5251    & 805.965\,x907 \\
  16&10317&-1.000\,002\,x545\,969\,5273    & 805.965\,x954 \\[1ex]
$\infty$&& -1.000\,002\,x545\,969\,5279(3) & 805.965\,x974(10)
  \end{tabular}
\end{ruledtabular}
\end{table}
For $r\leq 12.0$ 
results obtained with James-Coolidge basis are more accurate,
while for $r>12.0$ the ECA basis functions give more accurate energy. 
Extrapolated results for the whole BO potential
curve in the range $0.1 - 20$ au are presented in Table \ref{table3}.
{\squeezetable
\begin{table}[htb]\renewcommand{\arraystretch}{0.85}
\caption{\label{table3} Numerical values for BO potential at different
  internuclear distance $r$. Results are obtained by extrapolation to complete set
  of basis functions}
\begin{ruledtabular}
\begin{tabular*}
{1.00\textwidth}{w{1.2}w{2.20}@{\hfill}w{2.2}w{2.20}}
\cent{r/\mathrm{au}} & \cent{E(r)} &\cent{r/\mathrm{au}} & \cent{E(r)} \\
\hline
 0.10    & 7.127\,216\,731\,132\,0(55)             & 3.00    &-1.057\,326\,268\,872\,6617(70)  \\
 0.20    & 2.197\,803\,295\,226\,18(33)         & 3.10    &-1.051\,333\,772\,268\,0178(65)  \\
 0.30    & 0.619\,241\,659\,796\,226(60)         & 3.20    &-1.045\,799\,661\,432\,432(17)  \\
 0.40    &-0.120\,230\,341\,178\,823(15)        & 3.30    &-1.040\,717\,365\,351\,395(11)  \\
 0.50    &-0.526\,638\,758\,743\,001(11)        & 3.40    &-1.036\,075\,395\,190\,7599(41) \\
 0.60    &-0.769\,635\,429\,485\,9092(80)        & 3.50    &-1.031\,858\,084\,855\,0934(29) \\
 0.70    &-0.922\,027\,461\,527\,4636(26)       & 3.60    &-1.028\,046\,308\,379\,7348(90)  \\
 0.80    &-1.020\,056\,666\,360\,5151(4)        & 3.70    &-1.024\,618\,188\,410\,962(10)  \\
 0.90    &-1.083\,643\,239\,958\,8343(8)       & 3.80    &-1.021\,549\,795\,533\,649(12)  \\
 1.00    &-1.124\,539\,719\,546\,8709(7)       & 3.90    &-1.018\,815\,827\,696\,496(12)  \\
 1.10    &-1.150\,057\,367\,738\,5650(3)        & 4.00    &-1.016\,390\,252\,950\,6681(55)  \\
 1.20    &-1.164\,935\,243\,440\,3099(7)       & 4.20    &-1.012\,359\,959\,683\,166(11)  \\
 1.25    &-1.169\,419\,627\,390\,9022(7)       & 4.40    &-1.009\,256\,516\,261\,5862(34) \\
 1.30    &-1.172\,347\,149\,038\,0904(8)       & 4.60    &-1.006\,895\,223\,822\,7406(46) \\
 1.32    &-1.173\,138\,736\,333\,4793(8)       & 4.80    &-1.005\,116\,006\,100\,3838(50)  \\
 1.34    &-1.173\,734\,874\,958\,3451(3)        & 5.00    &-1.003\,785\,658\,583\,9706(38) \\
 1.36    &-1.174\,148\,498\,570\,4193(9)        & 5.20    &-1.002\,796\,816\,311\,2547(33) \\
 1.38    &-1.174\,391\,683\,632\,2532(6)       & 5.40    &-1.002\,065\,057\,209\,7059(27) \\
 1.39    &-1.174\,452\,917\,278\,4574(4)        & 5.60    &-1.001\,525\,251\,886\,6137(16) \\
 1.40    &-1.174\,475\,714\,220\,4434(5)       & 5.80    &-1.001\,127\,880\,852\,4173(36)  \\
 1.41    &-1.174\,461\,370\,870\,6800(11)        & 6.00    &-1.000\,835\,707\,655\,1804(23)  \\
 1.42    &-1.174\,411\,141\,239\,2317(11)       & 6.50    &-1.000\,400\,548\,534\,5376(60)   \\
 1.44    &-1.174\,207\,836\,585\,0950(11)       & 7.00    &-1.000\,197\,914\,480\,0381(38)  \\
 1.46    &-1.173\,875\,042\,749\,2034(8)       & 7.50    &-1.000\,102\,106\,147\,8089(22)  \\
 1.48    &-1.173\,421\,418\,292\,0817(12)       & 8.00    &-1.000\,055\,604\,973\,0730(4)  \\
 1.50    &-1.172\,855\,079\,578\,5838(12)       & 8.50    &-1.000\,032\,171\,832\,8288(55)  \\
 1.55    &-1.170\,994\,919\,897\,0180(6)       & 9.00    &-1.000\,019\,781\,832\,4911(2)  \\
 1.60    &-1.168\,583\,373\,371\,4593(8)       & 9.50    &-1.000\,012\,856\,876\,8268(2)  \\
 1.70    &-1.162\,458\,726\,898\,4588(5)       &10.00    &-1.000\,008\,755\,746\,0515(1)   \\
 1.80    &-1.155\,068\,737\,611\,6094(11)       &10.50    &-1.000\,006\,189\,995\,1069(1)   \\
 1.90    &-1.146\,850\,697\,029\,6887(28)       &11.00    &-1.000\,004\,505\,989\,4362(1)  \\
 2.00    &-1.138\,132\,957\,132\,6480(34)       &11.50    &-1.000\,003\,356\,174\,5754(1)  \\
 2.10    &-1.129\,163\,836\,101\,3193(40)        &12.00    &-1.000\,002\,545\,969\,5285(1)  \\
 2.20    &-1.120\,132\,116\,849\,2218(48)        &13.00    &-1.000\,001\,529\,286\,6698(1) \\
 2.30    &-1.111\,181\,765\,204\,4391(17)        &14.00    &-1.000\,000\,960\,680\,7911   \\
 2.40    &-1.102\,422\,606\,011\,326(65)         &15.00    &-1.000\,000\,625\,453\,6319 \\
 2.50    &-1.093\,938\,129\,955\,879(75)         &16.00    &-1.000\,000\,419\,586\,3122 \\
 2.60    &-1.085\,791\,237\,396\,1321(12)        &17.00    &-1.000\,000\,288\,826\,2392  \\
 2.70    &-1.078\,028\,484\,183\,8287(46)        &18.00    &-1.000\,000\,203\,340\,5059  \\
 2.80    &-1.070\,683\,233\,481\,4249(55)        &19.00    &-1.000\,000\,146\,028\,2368    \\
 2.90    &-1.063\,778\,008\,806\,0211(65)        &20.00    &-1.000\,000\,106\,740\,1283
\end{tabular*}
\end{ruledtabular}
\end{table}}
The ECA functions have the right large $r$ behavior
and thus can be used to study the long range tails in the interatomic
interactions. Particularly interesting is the long range asymptotics
of the exchange energy $\Delta E$, which is
the difference between singlet and triplet electronic energies.
Table \ref{table4} presents our numerical results for the exchange energy 
at internuclear distances $r=6$ -- $20$ au.  
\begin{table}[htb]\renewcommand{\arraystretch}{0.85}
\caption{\label{table4} Exchange energy $\Delta E = E_u-E_g$ at
  different internuclear distances}
\begin{ruledtabular}
  \begin{tabular}{c.}
 $r$ & \cent{e^{2\,r}\,r^{-5/2}\,\Delta E}  \\ \hline
  6 &  1.886\,x757\,524(7) \\
  7 &  1.798\,x767\,166(8) \\
  8 &  1.736\,x967\,949(9) \\
  9 &  1.692\,x575\,090(10) \\
  10 & 1.659\,x770\,272(12) \\
  11 & 1.634\,x919\,801(15) \\
  12 & 1.615\,x710\,667(20)  \\
  13 & 1.600\,x619\,338(31) \\
  14 & 1.588\,x607\,797(56) \\
  15 & 1.578\,x947\,36(11) \\
  16 & 1.571\,x113\,18(24) \\
  17 & 1.564\,x718\,78(50)  \\
  18 & 1.559\,x474\,1(11)   \\
  19 & 1.555\,x157\,8(22)   \\
  20 & 1.551\,x599\,0(44)     
  \end{tabular}
\end{ruledtabular}
\end{table}
It was found by Herring and Flicker in \cite{herring} that the asymptotic limit is
$e^{-2\,r}\,r^{5/2}\,\bigl(\gamma + O(r^{-1/2})\bigr)$ with $\gamma=1.636\,571\,460\,2...$. 
The fit to our numerical data, see Fig.~\ref{figure1}, gives $\gamma=1.6(1)$
what can be regarded as a first numerical confirmation of the Herring and
Flicker result.  When assuming their constant, the fit of first three terms
in $1/\sqrt{r}$ expansion gives:
$-{1.11732}/{\sqrt{r}} + {2.13187}/{r} + {5.169}/{r^{3/2}}$,
what indicates very slow convergence of this expansion at typical
interatomic distances $r\sim 10$.

\section{Summary}
We have demonstrated the applicability of analytic formulas for
two-center two-electron integrals in high precision calculations
of Born-Oppenheimer potential for the ground electronic state
of the hydrogen molecule. The symmetric James-Coolidge basis
with as much as 22 000 functions provided energies with relative precision 
of about $10^{-15}$ for internuclear distances up to 10 au.
The extended Heitler-London basis with about 10 000 functions
give similarly  accurate description
at  internuclear distances $r \sim 10$ au and greater.
The extension of the analytic approach to excited states of H$_2$ and other
diatomic molecules, such as HeH$^+$ requires the evaluation of Slater integrals
with arbitrary nonlinear parameters (Ko\l os-Wolniewicz basis \cite{kw_basis}). 
These integrals are solved analytically,
see Ref. \cite{rec_h2}, but the explicit expression for higher powers of electron 
distances is very lengthy. This may need further work to find a more compact 
analytical form. 

Further improvement in energy requires more accurate calculations 
of relativistic effects, possibly including nonadiabatic corrections.
For this, integrals with quadratic inverse powers of interparticle distances
have to be worked out analytically, and this is not a simple problem,
at least for arbitrary nonlinear parameters.

Even more important is the extension of this approach
to few and many electron diatomic systems. Integrals with
(uncorrelated) Slater functions with arbitrary nonlinear parameters
can be evaluated using recursion relations of Ref. \cite{rec_h2},
this includes also the exchange integrals involving Coulomb interaction
between electrons. More challenging would be implementation of 
coupled-clusters with the explicit $r_{12}$ factor. The analytic form for
the corresponding three-electron integrals is however not yet known.

\section*{Acknowledgments}
The author wishes to acknowledge interesting discussions with Bogumi\l\ Jeziorski
and Jacek Komasa. This work was supported by NIST
through Precision Measurement Grant PMG 60NANB7D6153.

\end{document}